\documentclass{ws-procs9x6}

\usepackage{slashed}

\begin{document}

\title{GAUGE NON-INVARIANCE AS TESTS OF EMERGENT GAUGE SYMMETRY}

\author{JOHN F. DONOGHUE$^{a*}$, MOHAMED M. ANBER$^b$ and UFUK AYDEMIR$^c$}

\address{Department of Physics, University of Massachusetts,\\
Amherst, MA 01003, USA\\
$^a$($^*$speaker) E-mail: donoghue@physics.umass.edu \\
$^b$ Email: manber@physics.umass.edu \\
$^c$ Email: uaydemir@physics.umass.edu}

\begin{abstract}
We motivate the concept of emergent gauge symmetry and discuss ways that this concept can be tested. The key idea is that if a symmetry is emergent, one should look for small violations of this symmetry because the underlying fundamental theory does not contain the symmetry. We describe our recent work implementing this idea in the gravity sector. We also describe the reasons why violations of gauge symmetry may well be linked to violations of Lorentz invariance.
\end{abstract}

\bodymatter

\section{Emergence and emergent symmetry}

The word ``emergence'' is appearing with increasing frequency in the particle physics literature. While it can mean different things to different people, at its heart it implies that properties and degrees of freedom that one observes differ drastically from the actual underlying physics. A classic example is the emergence of phonons in materials. The underlying physics is atoms bumping into each other, but the low energy description is that of waves that satisfy the massless wave equation. These waves can be quantized and used to explain the heat capacity of materials, for example, but at the smallest distance scales there in no such thing as a fundamental phonon field.

There is a familiar example of emergence in particle physics - that of pions and the chiral lagrangian. Here the underlying degrees of freedom are quarks and gluons with the interactions of QCD, yet the lightest fields are pions with a vastly different lagrangian,
\begin{equation}
{\cal L}_{QCD} = -\frac14 F^{\mu\nu}F_{\mu\nu} +\bar{\psi} i \slashed{D} \psi~~\to ~~{\cal L}_{eff} = \frac{F_\pi^2}{4}Tr [\partial_\mu U^\dag\partial^\mu U]
\end{equation}
with $U=\exp{(i\tau\cdot \pi/F_\pi)}$. If two of the quarks had been massless, the pions would also be massless and they would be active in atomic physics, well below the QCD scale. It would have taken an extraordinary leap of imagination for the atomic physicists of the 1930's to infer the underlying structure of QCD when confronted with these massless bosons.

However, our subject here is actually somewhat different - that of ``emergent symmetry''. The pion example above is {\em not} emergent symmetry. The chiral symmetry of the pion lagrangian is a reflection of the underlying chiral symmetry of QCD. However, symmetries can also emerge. As an example, consider the classic textbook derivation of the wave equation from masses interacting with their neighbors along a one-dimensional array. If one takes the general interaction potential $V(y_i -y_{i-1})$ near its minimum it is approximately harmonic $\sim \frac12 k(y_i -y_{i-1})^2$. Taking the continuum limit leads to the 2D wave equation for waves on a string, i.e. the massless Klein-Gordon equation.
\begin{equation}
S = \int dt\sum_i~[\frac12 m \dot{y}_i^2 -V(y_i -y_{i-1}) ] ~~\to~~\int d^2x \frac12  \partial_\mu \phi\partial^\mu\phi
\end{equation}
where here $\partial^\mu \equiv (\partial_t/v_s, \partial_x)$ with $v_s$ being the speed of sound. This latter form has symmetries that the original lagrangian did not have. The wave lagrangian is translation invariant, while the original one was not - this is a consequence of the continuum limit. There is also an emergent Lorentz-like symmetry - the wave lagrangian is invariant under Lorentz transformations, with $v_s$ in place of $c$ of course. There is also a shift symmetry $\phi \to \phi +c$, which keeps the wave massless, which is not really emergent. This is a reflection of an underlying symmetry of the original system of shifting all the $y_i$ by a constant, even though the wave field is not defined in the original lagrangian.

This example can also be used to illustrate the key principle of the phenomenology of emergent symmetry. Since the symmetry is not exact, one should look for evidence of the violation of the symmetry. In the ``waves on a string'' case, we can see this explicitly. If we look at the next term in the expansion of the potential about the minimum,
\begin{equation}
V(y_i -y_{i-1})=  \frac12 k(y_i -y_{i-1})^2 + \lambda (y_i -y_{i-1})^4 +....
\end{equation}
when we take the continuum limit, the differences turn into spatial derivatives and we arrive at the lagrangian
\begin{equation}
S = \int d^2x ~\left[\frac12  \partial_\mu \phi\partial^\mu\phi+ \bar{\lambda}(\frac{\partial \phi}{\partial x})^4\right]
\end{equation}
where the last term involves only spatial derivatives. This then breaks the emergent Lorentz-like symmetry.

This should be a general feature of emergent symmetry. If the low energy symmetry is not shared by the full theory, there will be some effects which do not have the symmetry. These symmetry violating terms are required to be suppressed if the symmetry has indeed emerged, but their presence is a indicator.

\section{Thinking about emergent symmetry}

The question then arises whether the symmetries of the Standard Model could be emergent. This of course would be a vastly different possibility than envisioned in the standard unification paradigm, where the low energy symmetries are part of larger symmetries at high energy.

There is no complete model for the emergence of the Standard Model symmetries. However, there has been a modest body of work - more than can be summarized accurately in the allotted write-up for this talk, so that space allows only the briefest of references\cite{emergence}.

Even without a complete model for emergent symmetries, there are aspects of the theory that we understand from other work.

On the positive side is the theorem by Deser\cite{Deser:1969wk} and others that says that if one has a massless spin-two field which couples to energy and momentum, including its own energy-momentum, the result of iterating the couplings will be general relativity. This is potentially an encouragement for emergence, as it implies that the full non-linear nature of general relativity could be the consequence of a seemingly simpler requirement of coupling the field to energy and momentum.

On the potentially negative side for emergence is the Weinberg-Witten theorem\cite{weinbergwitten}, which states that gravitons and Yang-Mills gauge bosons cannot be emergent from an underlying theory with Lorentz invariance. This follows from the requirement a Lorentz invariant emergent theory would lead to a Lorentz covariant energy-momentum tensor and gauge current. However the energy-momentum tensor of the physical gravitons and the gauge current of physical YM gauge bosons are not Lorentz covariant\cite{weinbergwitten} - although this feature is hidden by the ability to make a gauge transformation on the fields simultaneously with a Lorentz boost. The Weinberg-Witten theorem then appears to indicate that Lorentz invariance must also be emergent at the same time as gauge symmetry.

Another potential obstacle is the Nielsen-Ninomiya theorem\cite{Nielsen:1980rz} that forbids chiral fermions on a lattice. One might hope to avoid the Weinberg-Witten theorem by considering a discretetized spacetime, which would get around the Lorentz invariance problem. However, in this case, chiral fermions are problematic.

Also part of the emergence-Lorentz connection is the need for a universal limiting velocity (the speed of light) for all fields. In our experience with emergent fields, each carries its own velocity - phonons and magnons do not propagate at the same speed. However, to match our world all types of fields need to have the same limiting velocity - a stringent requirement.

\section{Phenomenology}

If a symmetry is emergent, it makes sense to look for potential small violations of that symmetry. Although much effort has gone into the study of the violation of discrete and global symmetries, very little work has been done on the violation of gauge symmetries.

Our published work in this area concerns gravity, in which case we are interested in studying the violation of general coordinate invariance\cite{Anber:2009qp}. Gravitational physics has the potential to be more sensitive to emergence than would usual gauge invariance tests. This is because we expect that the signal of emergence would be suppressed and gravity is itself suppressed by two inverse powers of the Planck mass. It is possible that a small signal could be relatively more visible in comparison to this already suppressed interaction. In addition, gravitational physics opens up vast time and distance scales, which could potentially be more revealing.

In our approach, we considered possible modifications to the Lagraingian  which contain two derivatives of the metric. This amounts to 5 possible terms expressed using the connection, of which one of which is ${\cal L}_3=-\,g^{\alpha\gamma}g^{\beta\rho}g_{\mu\nu}\Gamma^\mu_{\alpha\beta}\Gamma^\nu_{\gamma\rho}$. Because these are not generally covariant, these are only consistent in a particular set of coordinates, which then forms a constraint on allowable solutions. We studied the full set of Lagrangians at linear order, and there are constraints on the coefficients  of the various Lagrangian terms at the $10^{-3}$ level from the bending of light. There are also constraints on the sign of some combinations of coefficients from the perturbative stability of the graviton propagator.

We also took the Lagrangian ${\cal L}_3$ of the previous paragraph and performed the full matching to the PPN formalism\cite{Will:1993ns}. Our work cannot be done in the standard PPN gauge of that formalism, so we transformed the PPN gauge to the coordinates appropriate for our Lagrangian. The strongest constraint comes from the orbital polarization of binary pulsars and gives a constraint on the coefficient of ${\cal L}_3$ at the $10^{-20}$ level, when normalized in the same way as the Einstein-Hilbert action. This is quite strong, although without a fundamental underlying emergent theory the implication for the possibility of emergence is not clear.

We also are exploring the possibility that gauge symmetry violation may come along with the violation of Lorentz invariance\cite{gauge}. This connection is motivated by the Weinberg-Witten theorem - in emergent theories the energy scale of Lorentz violation and the scale of emergence could be the same. This connection may provide more sensitivity, because in many cases the tests of Lorentz invariance are quite strong and can be extended to gauge violating interactions.

There are many possible directions that one can consider in the study of emergent symmetry. While the unification paradigm has been pursued for several decades, the topic of emergent symmetry has been only lightly explored. That is part of the value of the subject - it provides a novel pathway for fundamental physics and one which may have new surprises. The possibility of testing the hypothesis exists, using probes which measure the violation of gauge symmetry and/or general covariance.

\section*{Acknowledgments}
This work has been supported in part by the NSF grants PHY- 055304 and PHY - 0855119, and in part by the Foundational Questions
Institute.

\end{document}